\documentclass[usenatbib]{mn2e}
\usepackage{amssymb}
\usepackage{graphicx}
\usepackage{color}

\newcommand{\be}{\begin{equation}}
\newcommand{\ee}{\end{equation}}
\newcommand{\nus}{\nu_{\mathrm{s}}}
\newcommand{\Lacc}{L_{\mathrm{acc}}}
\newcommand{\Lheat}{L_{\mathrm{heat}}}
\newcommand{\Qnuc}{Q_{\mathrm{nuc}}}
\newcommand{\Qgrav}{Q_{\mathrm{grav}}}
\newcommand{\mb}{m_{\mathrm{b}}}
\newcommand{\densnuc}{\rho_{\mathrm{nuc}}}
\newcommand{\Lnu}{L_\nu}
\newcommand{\Lmu}{\Lnu^{\mathrm{MU}}}
\newcommand{\Lsf}{\Lnu^{\mathrm{SF}}}
\newcommand{\Tc}{T}
\newcommand{\Tmu}{\Tc^{\mathrm{MU}}}
\newcommand{\Tcp}{T_{\mathrm{cp}}}
\newcommand{\Tcn}{T_{\mathrm{cn}}}
\newcommand{\Tcnmax}{T_{\mathrm{cn,max}}}
\newcommand{\dcnpeak}{\rho_{\mathrm{cn,peak}}}
\newcommand{\Lsflo}{\Lnu^{\mathrm{npSF}}}
\newcommand{\Lsfhi}{\Lnu^{\mathrm{pSF}}}
\newcommand{\Tsflo}{\Tc^{\mathrm{npSF}}}
\newcommand{\Tsfhi}{\Tc^{\mathrm{pSF}}}
\newcommand{\Lq}{L_{\mathrm{q}}}
\newcommand{\yign}{y_{\mathrm{ign}}}

\title[Superfluidity and NS core temperatures]{Superfluid effects on
gauging core temperatures of neutron stars in low-mass X-ray binaries}
\author[W.~C.~G. Ho]{Wynn C.~G. Ho$^1$\thanks{Email: wynnho@slac.stanford.edu}\\
$^1$School of Mathematics, University of Southampton, Southampton, SO17 1BJ}
\date{\today}

\begin{document}
\pagerange{\pageref{firstpage}--\pageref{lastpage}} \pubyear{2011}

\maketitle

\label{firstpage}

\begin{abstract}
Neutron stars accreting matter from low-mass binary companions are
observed to undergo bursts of X-rays due to the thermonuclear explosion
of material on the neutron star surface.
We use recent results on superfluid and superconducting properties
to show that the core temperature in these neutron stars may not be
uniquely determined for a range of observed accretion rates.
The degeneracy in inferred core temperatures could contribute to explaining
the difference between neutron stars which have very short recurrence times
between multiple bursts and those which have long recurrence times between
bursts:
short bursting sources have higher temperatures and normal neutrons in
the stellar core, while long bursting sources have lower temperatures and
superfluid neutrons.
If correct, measurements of the lowest luminosity from among the short
bursting sources and highest luminosity from among the long
bursting sources can be used to constrain the critical temperature
for the onset of neutron superfluidity.
\end{abstract}

\begin{keywords}
accretion, accretion discs ---
dense matter ---
neutrinos ---
stars: neutron ---
X-rays: binaries ---
X-rays: bursts
\end{keywords}

\section{Introduction} \label{sec:intro}

Neutron stars (NSs) are created in the collapse and subsequent supernova
explosion of massive stars.
NSs begin their lives very hot (with core temperatures
$k\Tc>10\mbox{ MeV}$) but cool rapidly by the emission of neutrinos.
Neutrino emission processes (and hence cooling of the NS) depend on
uncertain physics at the extreme supra-nuclear densities of the NS core
(see \citealt{tsuruta98,yakovlevpethick04,pageetal06}, for review).
Current theories indicate that the core may contain particles beyond
what makes up normal matter, such as hyperons and deconfined quarks,
and even the normal matter may be composed of a neutron superfluid and
proton superconductor
(\citealt{migdal59}; see, e.g., \citealt{lattimerprakash04,haenseletal07},
for review).

Superfluidity has two important effects on neutrino emission and NS cooling:
(1) suppression of emission mechanisms, like modified Urca processes,
that involve superfluid constituents and
(2) enhanced emission due to Cooper pairing of nucleons when the
temperature decreases just below a critical temperature
\citep{yakovlevpethick04,pageetal06}.
The measurement of rapid cooling of the
(youngest in our Galaxy; \citealt{hoheinke09})
NS in the Cassiopeia~A supernova remnant \citep{heinkeho10,shterninetal11}
provides the first direct evidence for the existence of superfluid
components in the core of a NS and constrains the critical
temperatures for the onset of superfluidity of neutrons $\Tcn$
(in the triplet state) and protons $\Tcp$ (in the singlet state),
i.e., $\Tcnmax\approx(5-9)\times 10^8$~K and $\Tcp\sim (2-3)\times 10^9$~K
\citep{pageetal11,shterninetal11}.

In contrast to the NS in Cassiopeia~A, many old NSs are found in binary
systems.  These binaries are seen in X-rays, which are produced when
material from the companion accretes onto the NS.
If the companion has a low stellar mass, the systems are known as
low-mass X-ray binaries (LMXBs).
Many LMXBs undergo bright X-ray bursts due to unstable thermonuclear burning
of hydrogen and/or helium in the surface layers of the NS
(see, e.g.,
\citealt{lewinetal93,bildsten98,strohmayerbildsten06,gallowayetal08},
for review).
Bursts are sometimes observed to recur in individual sources,
and recurrence times between multiple bursts span a wide range,
from minutes to days \citep{gallowayetal08,keeketal10}.
However, recurrence times $\lesssim 1\mbox{ hr}$ are too short for the
NS to accrete enough fuel for subsequent bursts
\citep{lewinetal93,woosleyetal04}.
\citet{keeketal10} studied the properties of these bursts and
found fifteen LMXBs that underwent short recurrence time X-ray bursts.
Only five of the NSs in these LMXBs have measured rotation rates $\nus$,
and all five possess $\nus\gtrsim 550\mbox{ Hz}$.
This would suggest that fast spins play an important role in causing short
recurrence times, e.g., through stellar r-mode oscillations \citep{hoetal11}.
However, the recent discovery of IGR~J17480$-$2446 (in the globular cluster
Terzan~5), which is a slow-spinning ($\nus=11\mbox{ Hz}$) LMXB
\citep{bordasetal10,strohmayermarkwardt10} that shows bursts with
recurrence times as short as $3\mbox{ min}$,
suggests rotation may not be a crucial ingredient \citep{mottaetal11}.

In light of the recent superfluid results from studies of the Cassiopeia~A NS,
we revisit the method used to infer core temperatures of NSs in LMXBs.
We demonstrate that superfluid effects can lead to non-uniqueness in the
inferred temperatures (if $\Tcn$ is low enough) and offer this high/low
temperature degeneracy as a possible explanation for the difference between
short/long recurrence time LMXBs.
In Section~\ref{sec:heat}, we briefly discuss the accretion-induced nuclear
heating of a NS core.
Section~\ref{sec:cool} describes the standard neutrino cooling mechanism,
as well as our model of superfluid neutrino cooling.
In Section~\ref{sec:temp}, we determine the core temperature for NSs in LMXBs
from a balance between heating and cooling and apply our results to short
and long recurrence time LMXBs.
We summarize in Section~\ref{sec:discuss}.

\vspace{-0.5cm}
\section{Accretion/Nuclear Heating} \label{sec:heat}

Many NSs in LMXBs are seen to undergo long-term accretion;
for each source, the measured average flux $F$ and distance $d$
provide an estimate of the accretion luminosity $\Lacc=4\pi d^2F$.
Accretion onto the NS surface releases a gravitational energy per nucleon
$\Qgrav=GM\mb/R=190\mbox{ MeV}(M/1.4\,M_\odot)(R/10\mbox{ km})^{-1}$,
where $M$ and $R$ are the NS mass and radius, respectively,
$\mb$ is the nucleon mass, and $M_\odot$ is the solar mass.
Only a small fraction of this energy diffuses into and heats the core
\citep{hanawafujimoto84,fujimotoetal87}.
On the other hand, compression by accreted matter induces nuclear reactions
in the deep crust, which release $\Qnuc\approx 1.5\mbox{ MeV nucleon$^{-1}$}$
\citep{haenselzdunik90,haenselzdunik08}, and this heats the core directly
by a luminosity \citep{brownetal98,brown00}
\be
\Lheat = (\Qnuc/\Qgrav)\Lacc = 0.0078\,\Lacc. \label{eq:lheat}
\ee
We note that r-mode heating yields $\Lheat=0.046(\nus/300\mbox{ Hz})\Lacc$
and is a more efficient heat source for fast-spinning NSs
\citep{brownushomirsky00,hoetal11}.

\vspace{-0.5cm}
\section{Neutrino Cooling} \label{sec:cool}

At high internal temperatures, neutron stars cool by neutrino emission
(see \citealt{tsuruta98,yakovlevpethick04,pageetal06}, for review);
cooling via neutrino emission dominates over photon emission
at our considered temperatures ($\Tc\gtrsim 10^8\mbox{ K}$).
The neutrino luminosity $\Lnu$ depends on the NS equation of state (EOS)
since $\Lnu$ is calculated by integrating the density-dependent emissivity
over the volume of the star.
Standard (slow) cooling is primarily determined by the modified Urca
processes.
For simplicity, we use the general result from \citet{shapiroteukolsky83}
\be
\Lmu\approx 7.4\times 10^{31}\mbox{ ergs s$^{-1}$ }(\Tc/10^8\mbox{ K})^8,
 \label{eq:lmu}
\ee
where we assume $M=1.4\,M_\odot$ and constant $\rho=\densnuc$ just
for the above equation.

\begin{figure}
\begin{center}
\resizebox{0.80\hsize}{!}{\includegraphics{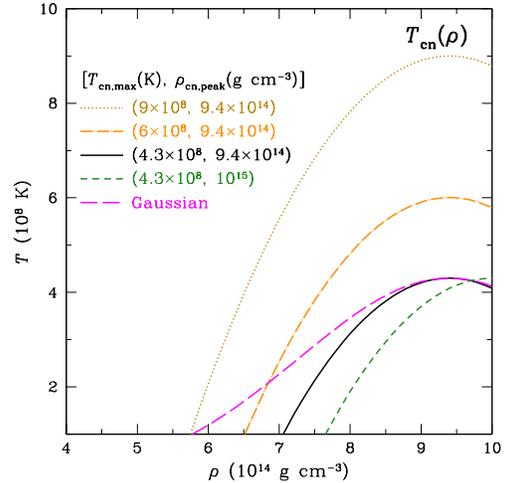}}
\caption{
Simple models of the neutron superfluid critical temperature $\Tcn(\rho)$
(see text for details).
See Fig.~\ref{fig:lnutcn} for the resulting neutrino luminosities using
these models.
}
\label{fig:tcn}
\end{center}
\vspace{-0.3cm}
\end{figure}

The modified Urca neutrino luminosity $\Lmu$ given by eq.~(\ref{eq:lmu})
assumes normal nucleons in the stellar core.
When nucleons are superfluid, the modified Urca processes involving these
nucleons are suppressed.  In addition, a new neutrino emission channel,
due to Cooper pair formation and breaking, opens up and is important near the
critical temperature for the onset of superfluidity
($\Tcp$ for protons and $\Tcn$ for neutrons;
see, e.g., \citealt{yakovlevetal99a}, and references therein).
We use \citet{yakovlevetal99,pageetal04,pageetal09} to calculate the
neutrino emissivities due to modified Urca processes,
accounting for superfluid suppression, and Cooper pair formation processes.
We take a constant $\Tcp=3\times 10^9$~K, so that
neutrino emission from Cooper pairing of protons is negligible, and
the only effect of superconducting protons is suppression of modified
Urca processes.
For $\Tcn(\rho)$, we use simplified models that depend on three parameters:
the maximum critical temperature $\Tcnmax$, the density at which this peak
occurs $\dcnpeak$, and the width of the peak.
For example, we consider a model that approximates model (a) of
\citet{shterninetal11}, i.e.,
\be
\Tcn(\rho)=\Tcnmax-6\times 10^7\mbox{ K}\left(\frac{\rho-\dcnpeak}{10^{14}
\mbox{ g cm$^{-3}$}}\right)^2. \label{eq:tcnquad}
\ee
Figure~\ref{fig:tcn} shows $\Tcn(\rho)$ for $\Tcnmax=4.3\times 10^8\mbox{ K}$
and $\dcnpeak=9.4\times 10^{14}\mbox{ g cm$^{-3}$}$.
Also shown is $\Tcn(\rho)$ for several other values of $\Tcnmax$ and
$\dcnpeak$.
The impact of the width of $\Tcn(\rho)$ is examined by comparing the quadratic
density-dependence of eq.~(\ref{eq:tcnquad}) with a $\Tcn(\rho)$ that is given
by a Gaussian profile
\be
\Tcn(\rho)=\Tcnmax\exp\left[-\left(
 \frac{\rho-\dcnpeak}{3\times 10^{14}\mbox{ g cm$^{-3}$}}\right)^2\right];
 \label{eq:tcngauss}
\ee
this is shown in Fig.~\ref{fig:tcn} for $\Tcnmax=4.3\times 10^8\mbox{ K}$ and
$\dcnpeak=9.4\times 10^{14}\mbox{ g cm$^{-3}$}$.
These phenomenological models for $\Tcn(\rho)$
are approximations of more detailed but uncertain theoretical models
(see, e.g., \citealt{lombardoschulze01,pageetal04,pageetal09}, and references
therein; see also \citealt{kaminkeretal02,gusakovetal04}).

The superfluid neutrino luminosity $\Lsf$ is obtained by integrating the
neutrino emissivities over the volume of the star and accounting for
$\Tcn(\rho)$.
We use a stellar model based on the Akmal-Pandharipande-Ravenhall EOS
\citep{akmaletal98,heiselberghjorthjensen99,gusakovetal05}
with $M=1.4M_\odot$ and $R=12$~km.
For simplicity, we have not taken into account other (less important)
neutrino emission processes in the core and crust,
nor have we examined the effect of the EOS, which determines the onset
of more efficient (fast) neutrino emission processes
(such as direct Urca or those associated with hyperon or quark condensates)
and partly determines the fraction of the NS that is superfluid;
these issues are beyond the scope of this work.

\vspace{-0.5cm}
\section{Neutron Star Core Temperatures} \label{sec:temp}

\begin{figure}
\begin{center}
\resizebox{0.85\hsize}{!}{\includegraphics{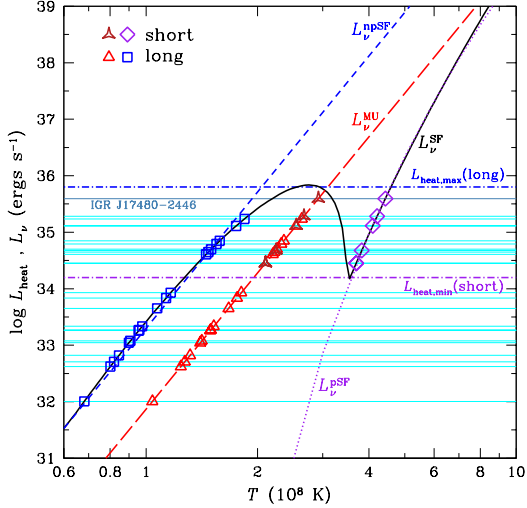}}
\caption{
Heat generated by accretion and nuclear reactions $\Lheat$ compared to the
neutrino cooling luminosity $\Lnu$ as a function of neutron star core
temperature $\Tc$.
The upper (lower) dot-dashed line is the highest (lowest) observed
$\Lheat$ from among all long (short) recurrence time bursts,
while the solid horizontal lines are $\Lheat$ for fast-spinning LMXBs and
IGR~J17480$-$2446.
The long-dashed line is the modified Urca luminosity $\Lmu$.
Triangles and starred-triangles indicate the intersection of $\Lheat$ and
$\Lmu$, which determines $\Tc$ for each long and short recurrence time LMXB,
respectively.
The thick solid line is $\Lsf$ with $\Tcnmax=4.3\times 10^8\mbox{ K}$ and
$\dcnpeak=9.4\times 10^{14}\mbox{ g cm$^{-3}$}$,
and squares and diamonds are where $\Lheat=\Lsf$ for each long and short LMXB,
respectively.
The short-dashed and dotted lines are approximate fits to $\Lsf$
in the strongly-superfluid and non-superfluid neutron regimes, respectively.
}
\label{fig:ltemp}
\end{center}
\vspace{-0.3cm}
\end{figure}

\begin{figure}
\begin{center}
\resizebox{0.85\hsize}{!}{\includegraphics{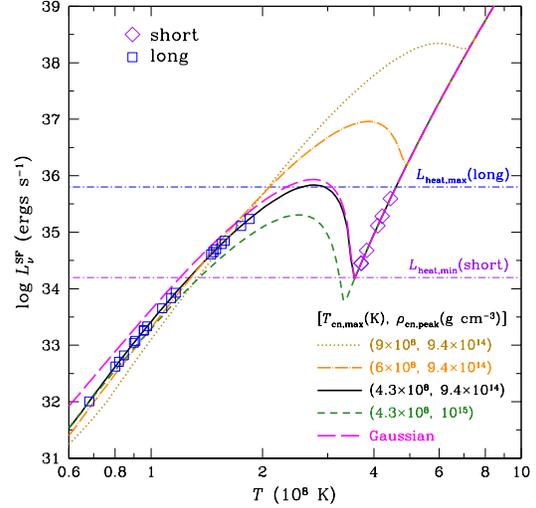}}
\caption{
Neutrino luminosity $\Lsf$ as a function of neutron star core
temperature $\Tc$, where the different curves are $\Lsf$ calculated
using models of the neutron superfluid critical temperature $\Tcn(\rho)$
shown in Fig.~\ref{fig:tcn}.
The upper (lower) horizontal dot-dashed line is the highest (lowest)
observed $\Lheat$ from among all long (short) recurrence time bursts.
Squares and diamonds are where
$\Lheat=\Lsf(4.3\times 10^8\mbox{ K},9.4\times 10^{14}\mbox{ g cm$^{-3}$})$
for each fast-spinning LMXB with long and short recurrence times, respectively.
}
\label{fig:lnutcn}
\end{center}
\vspace{-0.3cm}
\end{figure}

Figure~\ref{fig:ltemp} shows the heating rate $\Lheat$
[eq.~(\ref{eq:lheat})] and neutrino luminosities $\Lnu$ [$=\Lmu$ from
eq.~(\ref{eq:lmu}) and $\Lsf$ for $\Tcnmax=4.3\times 10^8\mbox{ K}$ and
$\dcnpeak=9.4\times 10^{14}\mbox{ g cm$^{-3}$}$].
The horizontal, solid lines indicate $\Lheat$ for LMXBs with
$\nus\gtrsim 100\mbox{ Hz}$, as well as IGR~J17480$-$2446 with
$\nus=11\mbox{ Hz}$, using fluxes and distances from
\citet{wattsetal08,ferrignoetal11,mottaetal11}.
Also plotted are $\Lheat$ calculated from the highest luminosity
($\approx 8\times 10^{37}\mbox{ ergs s$^{-1}$}$) seen amongst all
long recurrence time LMXBs and the lowest luminosity
($\approx 2\times 10^{36}\mbox{ ergs s$^{-1}$}$)
seen amongst all short recurrence time LMXBs \citep{keeketal10};
hereafter we follow \citet{keeketal10} in defining ``long'' LMXBs as
those that undergo multiple bursts with recurrence times
$\ge 1\mbox{ hr}$ and ``short'' LMXBs as those with recurrence times
$\le 1\mbox{ hr}$.  The various symbols denote $\Lheat=\Lnu(\Tc)$.
Figure~\ref{fig:lnutcn} compares $\Lsf$ for the different $\Tcn(\rho)$
profiles shown in Fig.~\ref{fig:tcn}.

{\it The intersection of the curves $\Lheat$ and $\Lnu$ yields the core
temperature of each NS}.
In the case of NSs cooling purely by modified Urca processes (unsuppressed
by nucleon superfluidity), the core temperature obtained from setting
$\Lheat=\Lmu$ is given by
\be
\Tmu = 1.3\times 10^8\mbox{ K }
 \left({\Lacc}/{10^{35}\mbox{ ergs s$^{-1}$}}\right)^{1/8}. \label{eq:tmu}
\ee
It is evident that, assuming the NS core is heated by accretion and cools
by modified Urca processes,
there is no clear distinction between NSs which have short and long recurrence
times: Fig.~\ref{fig:ltemp} shows the $\Tc$ of several long sources
interspersed between the $\Tc$ of short sources.

When superfluid effects are taken into account, there are pronounced
differences.
At $\Tc\gtrsim\Tcnmax$ (and $\Tc\ll\Tcp$), neutrons are normal while
protons are superconducting.
The latter suppresses modified Urca processes, so that $\Lsf$ ($<\Lmu$)
is approximately given by (see Fig.~\ref{fig:ltemp})
\be
\Lsfhi\approx 4\times 10^{39}\mbox{ ergs s$^{-1}$}[\log(T/10^8\mbox{ K})]^{21}.
\label{eq:lsfhi}
\ee
Setting $\Lsfhi=\Lheat$, the core temperature is then
\be
\log\Tsfhi = 8+0.5\left(\Lacc/10^{35}\mbox{ ergs s$^{-1}$}\right)^{1/21}.
 \label{eq:tsfhi}
\ee
Since cooling is less efficient in this case, the inferred core temperatures
are higher than those obtained from eq.~(\ref{eq:tmu}).
At $\Tc<\Tcnmax$, neutrino emission is enhanced due to the formation of
neutron Cooper pairs; this new emission channel dominates the modified
Urca emission.
Cooling is more efficient and can result in a lower inferred $\Tc$.
When neutrons (and protons) are strongly superfluid ($\Tc\ll\Tcnmax$),
\citet{gusakovetal04} find that the neutrino luminosity from Cooper pair
formation has the same $\Tc^8$-dependence as the modified Urca processes
but with a much higher efficiency
[i.e., larger coefficient in eq.~(\ref{eq:lmu})].
We find $\Lsf$ can be fit by (see Fig.~\ref{fig:ltemp})
\be
\Lsflo \sim (20-30)\Lmu; \label{eq:lsflo}
\ee
from $\Lsflo=\Lheat$, we obtain
\be
\Tsflo = 9\times 10^7\mbox{ K }
 \left(\Lacc/10^{35}\mbox{ ergs s$^{-1}$}\right)^{1/8}. \label{eq:tsflo}
\ee

The core temperature of NSs in relatively high-luminosity LMXBs may not be
uniquely determined.
If $\Tcnmax\lesssim 8\times 10^8\mbox{ K}$
(see Figs.~\ref{fig:ltemp} and \ref{fig:lnutcn}),
there are two thermally stable (see below) values of the core temperature
associated with a single observed accretion luminosity,
for a range of $\Lsf=\Lheat$.
For example, there is a factor of $< 3$ difference in the
inferred $\Tc$ if $\Lacc\sim (0.2-9)\times 10^{37}\mbox{ ergs s$^{-1}$}$
and $\Tcnmax=4.3\times 10^8$~K.
The persistent luminosities of all LMXBs that show short recurrence time
bursts lie within this range \citep{keeketal10}.
To highlight this point, we place the six short LMXBs with measured spin
periods ($\Lacc$ of the LMXB Aql~X-1 and EXO~0748$-$676 are very similar
and thus their inferred $\Tc$ are not noticeably different)
on the high-temperature $\Lsfhi$-branch
and the seven long LMXBs on the low-temperature $\Lsflo$-branch.
Note that there can be three values of $\Tc$ that cross each horizontal
$\Lheat$;
however the intermediate temperature is thermally unstable since a decrease
in temperature leads to an increase in neutrino luminosity and hence
causes even more rapid cooling.

At present, it is not known what causes LMXBs to undergo short versus long
recurrence time bursts.
Possibilities include variations in fraction of fuel burnt, mass accretion
rate, or composition of accreted matter.
As we have shown above, if $\Tcnmax\sim (4-5)\times 10^8$~K
[note the rapid cooling of the Cassiopeia~A NS indicates
$\Tcnmax\approx (5-9)\times 10^8$~K; \citealt{pageetal11,shterninetal11}],
another possibility is that short LMXBs have intrinsically hotter core
temperatures than long LMXBs.
This would indicate that neutrons are normal in the core of short recurrence
time LMXBs (so that $\Lsf$ is given by $\Lsfhi$),
while core neutrons are superfluid in the long recurrence time LMXBs
(so that $\Lsf\sim\Lsflo$).

A core temperature that is a factor of three hotter produces a surface
temperature that is $\sim 3^{1/2}$ hotter \citep{gudmundssonetal82} and a
surface flux that is $\sim (3^{1/2})^4$ brighter.
The higher surface temperature and flux may be sufficient to alter the
temperature in the nuclear burning layers and shorten the time intervals
between ignition of unstable burning.
For matter accreting at a rate $\dot{M}$, the time to replenish the nuclear
fuel (i.e., burst recurrence time) is $\sim R^2\yign/\dot{M}\propto T^{-5/2}$,
where $\yign$ is the ignition depth/column
(\citealt{bildsten98}; see also \citealt{cummingbildsten00}).
Thus a factor of three higher core temperature could shorten the recurrence
time by at least that amount.
As discussed in Sec.~\ref{sec:intro}, previous works calculate burst
recurrence times that are longer than observed, but numerical simulations
currently being performed suggest shorter times may be possible
(Keek \& Heger, in preparation).

\begin{figure}
\begin{center}
\resizebox{0.83\hsize}{!}{\includegraphics{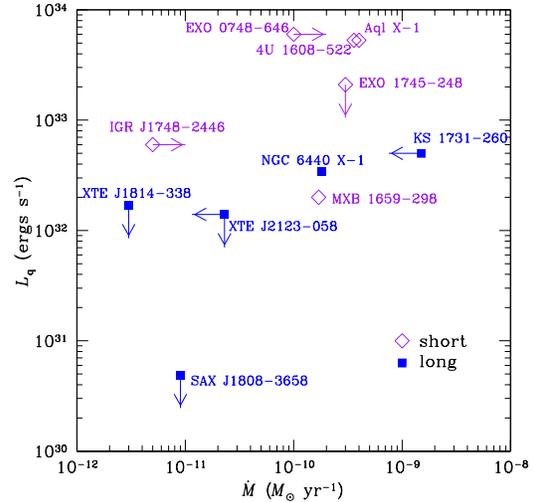}}
\caption{
Quiescent luminosity $\Lq$ as a function of time-averaged mass accretion rate
$\dot{M}$ of LMXBs that are seen to produce multiple nuclear X-ray bursts.
Diamonds and squares are for short and long recurrence time LMXBs,
respectively, and arrows indicate upper or lower limits.
}
\label{fig:lq}
\end{center}
\vspace{-0.3cm}
\end{figure}

A consequence of (possibly) higher core temperatures in short LMXBs is
that one might expect the quiescent luminosity $\Lq$
(i.e., when the NS is not accreting significantly)
of short LMXBs to be higher than that of (cooler) long LMXBs.
Previous works (see, e.g., \citealt{colpietal01,yakovlevetal04};
see however \citealt{levenfishhaensel07}) studying quiescent emission
do not account for superfluidity in the appropriate regime and thus do not
see the effects described here.
Figure~\ref{fig:lq} shows $\Lq$ for LMXBs with measured nuclear X-ray bursts
\citep{heinkeetal07,heinkeetal09,gallowayetal08,degenaarwijnands11,degenaaretal11,diaztrigoetal11}.
Note that surface burning effects can dominate core temperature variations
at instantaneous $\dot{M}\gtrsim 6\times 10^{-9}\,M_\odot\mbox{ yr$^{-1}$}$.
Note also that \citet{levenfishhaensel07} studied predictions for $\Lq$
and $\dot{M}$ and found that neutron superfluidity can create a dichotomy
amongst LMXBs.
Though there are many uncertainties involved, especially in distance and
accretion rates,
the observations suggest that short recurrence time LMXBs may be
intrinsically hotter.

Finally, if short recurrence time LMXBs do indeed possess hotter core
temperatures, then measurements of the minimum and maximum accretion
luminosities of bursts from short LMXBs and long LMXBs, respectively,
can be used to constrain the neutron superfluid critical temperature
$\Tcn(\rho)$.
This is illustrated in Fig.~\ref{fig:lnutcn}, where it is clear that the
accretion luminosities for LMXBs can constrain $\Tcnmax$ and $\dcnpeak$,
while the width of $\Tcn(\rho)$ is not as important in determining the
qualitative behavior of $\Lsf$.

\vspace{-0.5cm}
\section{Discussion} \label{sec:discuss}

We used the accretion luminosity measured from observations of LMXBs
to determine the heating rate of the NSs in these systems.
By balancing heating with cooling (via neutrino emission), we determined
NS core temperatures.
Uncertainties in the accretion/heating efficiency and nuclear energy
release have a small effect on the inferred temperatures because of the
strong temperature scalings in the neutrino emissivities.
We found that neutrino emission from Cooper pairing neutrons can yield a
non-unique determination of the core temperature.
We explored one possible implication, i.e.,
the observed variation in recurrence times between multiple
nuclear X-ray bursts could be a manifestation of differences
(by a factor of $\lesssim 3$) in NS core temperature.
LMXBs that undergo nuclear-powered bursts with long recurrence times
have lower core temperatures and neutrons that are superfluid,
while those with short recurrence time bursts have higher core
temperatures and normal neutrons.
Thus LMXBs which experience multiple bursts could provide constraints
on properties of neutron superfluidity.
We note that we have not examined the effect of higher temperatures on
different burning regimes and implications for, e.g., long bursts
and superbursts (see, e.g., \citealt{cummingetal06,strohmayerbildsten06},
and references therein).

Previous studies find that sequences of short recurrence bursts
involves, at most, a quadrupole set of bursts \citep{keeketal10}.
However, the recently discovered source, IGR~J17480$-$2446,
shows tens of bursts in a single event \citep{mottaetal11}.
This could be the result of IGR~J17480$-$2446 being the hottest of the
known LMXBs (see Fig.~\ref{fig:ltemp}).
Note though that the short recurrence time bursts from IGR~J17480$-$2446
may be different in nature than those seen in other sources
(L. Keek, private comm.).

A natural question is what determines the state of the neutrons,
or alternatively, which neutrino luminosity branch
(see Figs.~\ref{fig:ltemp} or \ref{fig:lnutcn}) does a particular LMXB lie on.
Presumably a young NS or one that sustains long-term heating of its core
above the peak in $\Lsf$ will be on the higher $\Tc$ (or $\Lsfhi$) branch.
If subsequent accretion initiates multiple bursts, these bursts will recur
on short timescales ($\lesssim 1\mbox{ hr}$).
If accretion does not significantly heat the core, then the NS will move
rapidly through the thermally unstable branch
(where the neutrino luminosity increases as the temperature decreases)
and shift to the lower $\Tc$ (or $\sim\Lsflo$) branch;
bursts from these NSs will recur with long timescales.
On the other hand, if the maximum critical temperature for neutron (triplet)
superfluidity $\Tcnmax\gtrsim 6\times 10^8$~K, then all LMXBs have core
temperatures given by $\Tsflo$ [see eq.~(\ref{eq:tsflo})].

\vspace{-0.5cm}
\section*{acknowledgments}
WCGH thanks Laurens Keek and Tom Maccarone for discussions, Peter Shternin
for providing the APR EOS, and the referee for valuable comments on nuclear
bursts.
WCGH appreciates the use of the computer facilities at KIPAC.
WCGH acknowledges support from STFC in the UK.

\vspace{-0.5cm}
\bibliographystyle{mnras}

\label{lastpage}

\end{document}